\documentclass[aps,prb,preprint,showpacs]{revtex4}
\usepackage{natbib}
\usepackage{epsfig}
\usepackage{color}

\begin{document}

\title{Jordan-Wigner transformation constructed for spinful fermions at
spin-1/2 in two dimensions.}
\author{Zsolt~Gulacsi} 
\affiliation{Department of Theoretical Physics, University of
Debrecen, H-4010 Debrecen, Bem ter 18/B, Hungary}
\date{\today }

\begin{abstract}
  Recently a Jordan-Wigner transformation was constructed for spinful fermions
  at S=1/2 spins in one dimension connecting the spin-1/2 operators to genuine
  spinful canonical Fermi operators. In the presented paper this exact
  transformation is generalized to two dimensions.
\end{abstract}

\maketitle


\section{Introduction}

The Jordan-Wigner transformation mapping the quantum spin-1/2 operators to
spinless Fermi operators in 1D represents a basic tool in theoretical physics
for almost one century \cite{R1}. After almost sixty years, maintaining the
quantum spin-1/2 to spinless fermion operator interconnection, it has been
generalized to two
dimensions \cite{NR2,NR3,NR4}, and at the beginning of the second millennium
also extended to $S > 1/2$ case \cite{NR5}. However, the spinless fermionic
nature of the used Fermi operators in this transformation hedges the mapping
possibilities of this transformation technique in between spin models (systems)
and realistic fermionic systems. For example, explicitly spin dependent
processes at fermionic level (as e.g. the many-body spin-orbit interaction)
cannot be involved.

As Ref.[\cite{NR6}] shows, the Jordan-Wigner transformation
can be used
for deducing exact solutions for 1D spin models by transforming these models in
spinless fermion models. But spinless fermion is in fact a mathematical
abstraction. The real fermions are spinful. Hence, the transformation of spin
operators to spinful Fermi operators provides the possibility to observe (find,
deduce)
equivalences in between spin models (systems) and realistic fermionic models
(systems). These equivalences provide new information for both models and could
transpose even approximated knowledge from one model to another. In the context
of 2D, where the description possibilities are much harder, such supplementary
information are clearly useful. These mapping possibilities were unknown
previously, and started to be known given by the presented transformation.
  
Recently a Jordan-Wigner type of transformation has been worked out
which maps in 1D the quantum spin-1/2 spin operators to genuine spinful
canonical Fermi operators \cite{NR1}. This transformation is extended here in
exact terms to two dimensions.

The main results of the paper are contained in Eqs.(\ref{NE9}, \ref{NE16},
\ref{NE17}, and \ref{NE30}). Based on these relations, the general 2D quantum
Heisenberg Hamiltonian in (\ref{NE8}) is exactly and explicitly transformed in
the language of spinful canonical Fermi operators in Eqs.(\ref{NE31}-
\ref{NE37}).

The remaining part of the paper is structured as follows: Section II shortly
recapitulates the results related to the 1D case, Section III describes the
transformation itself in 2D, Section IV explicitly transforms a 2D general
Heisenberg Hamiltonian in spinful canonical Fermi operator image, and Section
V containing the summary and discussions closes the presentation.

\section{Summary of the D=1 results.}

The spinful Jordan-Wigner transformation for $S=1/2$ in 1D has been defined
via \cite{NR1}
\begin{eqnarray}
&&\hat S^x_i = \frac{\hat a^{\dagger}_i + \hat a_i}{2X} + \frac{\hat b^{\dagger}_i +
\hat b_i}{2Y}, \quad \hat S^y_i = \frac{\hat a^{\dagger}_i - \hat a_i}{2iZ} +
\frac{\hat b^{\dagger}_i - \hat b_i}{2iW},
\nonumber\\
&&\hat S^z_i = - \frac{1}{2} ( \frac{1}{XZ} + \frac{1}{YW} ) +
(\frac{\hat a^{\dagger}_i \hat a_i}{XZ} +\frac{\hat b^{\dagger}_i \hat b_i}{YW}
) + \frac{1}{2} (\frac{1}{YZ} - \frac{1}{XW} ) (\hat a^{\dagger}_i
\hat b^{\dagger}_i + \hat b_i \hat a_i ) 
\nonumber\\
&& \hspace*{0.6cm} + \: \: \frac{1}{2} (\frac{1}{YZ}+ 
\frac{1}{XW})(\hat a^{\dagger}_i \hat b_i + \hat b^{\dagger}_i \hat a_i ), 
\label{NE1}
\end{eqnarray}
where the $\hat a_i, \hat b_i$ operators anticommute on-site, and commute
inter-sites, and the scalars $X,Y,Z,W$ satisfy the requirements $1=\frac{1}{X^2}
+\frac{1}{Y^2}$, $1=\frac{1}{Z^2}+\frac{1}{W^2}$ (so one has two free parameters
in the transformation). Please note that instead of $\hat S^x_i,
\hat S^y_i$ in the first row
of  Eq.(\ref{NE1}), we can use $\hat S^{+}_i, \hat S^{-}_i$ as well, where
$\hat S^{\pm}_i = \hat S^x_i \pm i \hat S^y_i = 1/2 [ \hat a^{\dagger}_i (1/X \pm
1/Z) + \hat a_i (1/X \mp 1/Z) + \hat b^{\dagger}_i (1/Y \pm 1/W) + \hat b_i
(1/Y \mp 1/W) ]$ holds.

In explaining these relations, I note that in this Eq.(\ref{NE1}),
$\hat S_i^x$, and $\hat S_i^y$ represent the starting point.These starting
expressions take into consideration that $\hat S_i^x, \hat S_i^y$ must be
Hermitian. Now, $\hat S_i^z$ is deduced from the $[\hat S_i^x,\hat S_i^y]=i
\hat S_i^z$ commutation relation prescribed for spin operators, where
$\hat S_i^z$ is also Hermitian. This required commutation relation provides
the $\hat S_i^z$ expression presented in Eq.(\ref{NE1}). But besides the
presented commutation relation, also $[\hat S_i^y,\hat S_i^z]=i \hat S_i^x$,
$[\hat S_i^z, \hat S_i^x]=i \hat S_i^y$, and $[\hat S^2,\hat S_i^{\alpha}]=0$,
where $\alpha=x,y,z$ must be satisfied. It can be easily checked that starting
from Eq.(\ref{NE1}), and taking into account the conditions presented
below Eq.(\ref{NE1}), namely $1/X^2 + 1/Y^2=1, 1/Z^2 + 1/W^2 =1$, all required
commutation relations are satisfied, and supplementary $\hat S^2=3/4$ (i.e.
S=1/2) holds. Excepting the presented two conditions for X,Y,Z,W, these
parameters are arbitrary.

Now, introducing the extended Jordan-Wigner type of transformations 
\begin{eqnarray}
&&\hat a_i = \exp [-i\pi \sum_{j=1}^{i-1}(\hat c^{\dagger}_j \hat c_j +
\hat f^{\dagger}_j \hat f_j)] \hat c_i, \quad
\hat b_i = \exp [-i\pi \sum_{j=1}^{i-1}(\hat c^{\dagger}_j \hat c_j +
\hat f^{\dagger}_j \hat f_j)] \hat f_i,
\nonumber\\
&&\hat a^{\dagger}_i = \hat c^{\dagger}_i \exp [i\pi \sum_{j=1}^{i-1}(
\hat c^{\dagger}_j \hat c_j + \hat f^{\dagger}_j \hat f_j)], \quad
\hat b^{\dagger}_i = \hat f^{\dagger}_i \exp [i\pi \sum_{j=1}^{i-1}(\hat c^{\dagger}_j
\hat c_j + \hat f^{\dagger}_j \hat f_j)],
\label{NE2}
\end{eqnarray}
where the $\hat c_i,\hat f_i$ are canonical Fermi operators, and considering
\cite{NR1}
$\hat c_i=\hat c_{i,\sigma}$, $\hat f_i =\hat c_{i,-\sigma}$, the mapping
(\ref{NE1})
transforms the spin-1/2  operators in genuine spinful canonical Fermi
operators.

Transforming the spin-1/2 operators, the described mapping
transforms in spinful fermionic language all Hamiltonians constructed from
these spin operators. For exemplification, below I analyze the case of the
Heisenberg Hamiltonians.

The 1D Heisenberg Hamiltonian
\begin{eqnarray}
\hat H = \sum_i [(J_x \hat S^x_i \hat S^x_{i+1} + J_y \hat S^y_i \hat S^y_{i+1}) +
J_z \hat S^z_i \hat S^z_{i+1}] = H_{xy} + H_z,
\label{NE3}
\end{eqnarray}
where one considers $J_x=J(1+\Gamma)$, $J_y=J(1-\Gamma)$, transforms as follows:
First we introduce the coupling constants
\begin{eqnarray}
&& K_{\sigma} = \frac{J}{4} [ (\frac{1}{X^2}-\frac{1}{Z^2}) + \Gamma
(\frac{1}{X^2}+\frac{1}{Z^2})],
\nonumber\\
&& K_{-\sigma} = \frac{J}{4} [ (\frac{1}{Y^2}-\frac{1}{W^2}) + \Gamma
(\frac{1}{Y^2}+\frac{1}{W^2})] = \frac{J}{4} [-(\frac{1}{X^2}-\frac{1}{Z^2}) +
\Gamma (2-(\frac{1}{X^2}+\frac{1}{Z^2}))],
\nonumber\\
&& t_{\sigma} = \frac{J}{4} [ (\frac{1}{X^2}+\frac{1}{Z^2}) + \Gamma
(\frac{1}{X^2}-\frac{1}{Z^2})],
\nonumber\\ 
&& t_{-\sigma} = \frac{J}{4} [ (\frac{1}{Y^2}+\frac{1}{W^2}) + \Gamma
(\frac{1}{Y^2}-\frac{1}{W^2})] =
\frac{J}{4} [(2- (\frac{1}{X^2}+\frac{1}{Z^2}) - \Gamma
(\frac{1}{X^2}-\frac{1}{Z^2})],
\nonumber\\
&& K = \frac{J}{4} [ (\frac{1}{XY}-\frac{1}{WZ}) + \Gamma
(\frac{1}{XY}+\frac{1}{WZ})] = \frac{J}{4} [ A + \Gamma B],
\nonumber\\
&& t = \frac{J}{4} [ (\frac{1}{XY}+\frac{1}{WZ}) + \Gamma
(\frac{1}{XY}-\frac{1}{WZ})] = \frac{J}{4} [ B + \Gamma A],
\label{NE4}
\end{eqnarray}
where $AB= \frac{1}{X^2}(1- \frac{1}{X^2}) -\frac{1}{Z^2}(1- \frac{1}{Z^2})$
and $A= \frac{1}{XY}-\frac{1}{WZ}$, $\frac{1}{Y^2}=1-\frac{1}{X^2}$,
$\frac{1}{W^2}=1-\frac{1}{Z^2}$. Note that the parameters $X,Z$ are
independent, and arbitrary.

The $H_{xy}$ Hamiltonian now becomes
\begin{eqnarray}
\hat H_{xy} &=& \sum_i \sum_{\sigma} [ K_{\sigma} \hat c^{\dagger}_{i,\sigma}
\hat c^{\dagger}_{i+1,\sigma}(1-2 \hat n_{i,-\sigma}) + t_{\sigma} \hat c^{\dagger}_{i,
\sigma} \hat c_{i+1,\sigma}(1-2 \hat n_{i,-\sigma}) + K \hat c^{\dagger}_{i,\sigma}
\hat c^{\dagger}_{i+1,-\sigma}(1-2 \hat n_{i,-\sigma})
\nonumber\\
&+& t \hat c^{\dagger}_{i,\sigma}
\hat c_{i+1,-\sigma}(1-2 \hat n_{i,-\sigma}) + h.c.]
\label{NE5}
\end{eqnarray}

Similarly, for $\hat H_z$ one finds
\begin{eqnarray}
\hat H_z = J_z \sum_i \{ \hat P_1(i,i+1) + \hat P_2(i,i+1) + \hat P_3(i,i+1) \} 
\label{NE6}
\end{eqnarray}
where, introducing the particle number operator $\hat n_{i,\sigma}=
\hat c^{\dagger}_{i,\sigma} \hat c_{i,\sigma}$, one has
\begin{eqnarray}
&&\hat P_1(i,i+1) = \frac{1}{4}(\frac{1}{XZ}+ \frac{1}{YW})^2 - \frac{1}{2}
(\frac{1}{XZ}+ \frac{1}{YW})\{ [ (\frac{\hat n_{i,\uparrow}}{XZ} +
\frac{\hat n_{i,\downarrow}}{YW}) + \frac{1}{2}(\frac{1}{YZ}- \frac{1}{XW})
(\hat c^{\dagger}_{i,\uparrow} \hat c^{\dagger}_{i,\downarrow} + \hat c_{i,\downarrow}
\hat c_{i,\uparrow})
\nonumber\\
&&+\frac{1}{2}(\frac{1}{YZ}+ \frac{1}{XW}) (\hat c^{\dagger}_{
i,\uparrow} \hat c_{i,\downarrow} + \hat c^{\dagger}_{i,\downarrow}\hat c_{i,\uparrow})] +
[ (\frac{\hat n_{i+1,\uparrow}}{XZ} + \frac{\hat n_{i+1,\downarrow}}{YW}) +
\frac{1}{2}(\frac{1}{YZ}- \frac{1}{XW})(\hat c^{\dagger}_{i+1,\uparrow}
\hat c^{\dagger}_{i+1,\downarrow} + \hat c_{i+1,\downarrow}\hat c_{i+1,\uparrow})
\nonumber\\
&&+\frac{1}{2}(\frac{1}{YZ}+ \frac{1}{XW}) (\hat c^{\dagger}_{i+1,\uparrow}
\hat c_{i+1,\downarrow} + \hat c^{\dagger}_{i+1,\downarrow}\hat c_{i+1,\uparrow})]\},
\nonumber\\
&&\hat P_2(i,i+1) =(\frac{\hat n_{i,\uparrow}}{XZ} +
\frac{\hat n_{i,\downarrow}}{YW})(\frac{\hat n_{i+1,\uparrow}}{XZ} +
\frac{\hat n_{i+1,\downarrow}}{YW}) + (\frac{\hat n_{i,\uparrow}}{XZ} +
\frac{\hat n_{i,\downarrow}}{YW}) [ \frac{1}{2}(\frac{1}{YZ}- \frac{1}{XW})
(\hat c^{\dagger}_{i+1,\uparrow}\hat c^{\dagger}_{i+1,\downarrow} + \hat c_{i+1,\downarrow}
\hat c_{i+1,\uparrow})
\nonumber\\
&&+ \frac{1}{2}(\frac{1}{YZ}+ \frac{1}{XW})
(\hat c^{\dagger}_{i+1,\uparrow}\hat c_{i+1,\downarrow} + \hat c^{\dagger}_{i+1,\downarrow}
\hat c_{i+1,\uparrow})] + (\frac{\hat n_{i+1,\uparrow}}{XZ} +
\frac{\hat n_{i+1,\downarrow}}{YW}) [  \frac{1}{2}(\frac{1}{YZ}- \frac{1}{XW})
(\hat c^{\dagger}_{i,\uparrow}\hat c^{\dagger}_{i,\downarrow} + \hat c_{i,\downarrow}
\hat c_{i,\uparrow})
\nonumber\\
&&+ \frac{1}{2}(\frac{1}{YZ}+ \frac{1}{XW})
(\hat c^{\dagger}_{i,\uparrow}\hat c_{i,\downarrow} + \hat c^{\dagger}_{i,\downarrow}
\hat c_{i,\uparrow})],
\nonumber\\
&&\hat P_3(i,i+1) = \frac{1}{4}(\hat c^{\dagger}_{i,\uparrow}\hat c^{\dagger}_{i,
\downarrow} + \hat c_{i,\downarrow} \hat c_{i,\uparrow})[(\frac{1}{YZ}-
\frac{1}{XW})^2 (\hat c^{\dagger}_{i+1,\uparrow}\hat c^{\dagger}_{i+1,\downarrow} +
\hat c_{i+1,\downarrow} \hat c_{i+1,\uparrow})
\nonumber\\
&&+ (\frac{1}{(YZ)^2} - \frac{1}{(XW)^2})
(\hat c^{\dagger}_{i+1,\uparrow} \hat c_{i+1,\downarrow} + \hat c^{\dagger}_{i+1,\downarrow}
\hat c_{i+1,\uparrow})] + \frac{1}{4}(\hat c^{\dagger}_{i,\uparrow}\hat c_{i,
\downarrow} + \hat c^{\dagger}_{i,\downarrow} \hat c_{i,\uparrow}) [ (\frac{1}{(YZ)^2}-
\frac{1}{(XW)^2})
\nonumber\\
&&*(\hat c^{\dagger}_{i+1,\uparrow} \hat c^{\dagger}_{i+1,\downarrow} +
\hat c_{i+1,\downarrow} \hat c_{i+1,\uparrow}) + (\frac{1}{YZ} + \frac{1}{XW})^2
(\hat c^{\dagger}_{i+1,\uparrow} \hat c_{i+1,\downarrow} + \hat c^{\dagger}_{i+1,\downarrow}
\hat c_{i+1,\uparrow})].
\label{NE7}
\end{eqnarray}

In order to obtain Eq.(\ref{NE7}), we introduce in Eq.(\ref{NE3})
the $\hat S_i^{\alpha}$ values from Eq.(\ref{NE1}), effectuate the required
products ($\hat S_i^{\alpha} \hat S_{i+1}^{\alpha}$) for all $\alpha=x,y,z$,
add the three $\alpha$ terms and obtain the Hamiltonian in term of
$\hat a_i$ and $\hat b_i$ operators. After this step, using Eq.(\ref{NE2}),
where $\hat c_i= \hat c_{i,\sigma}, \hat f_i=\hat c_{i,-\sigma}$, one obtains
Eq.(\ref{NE7}). Note that the $X,Y,Z,W$ parameters are not unknown quantities.
They can be such chosen to map a fermionic model of interest given in terms of
realistic (spinful) fermions in a quantum spin model language written for
S=1/2 spins.

\section{
The D=2 dimensional spinful Jordan-Wigner
transformation.
}

One considers a general D=2 dimensional lattice, with Bravais vectors ${\bf
a_1}, {\bf a_2}$  (say for simplicity, in x and y directions). An arbitrary
lattice
site n is given by a pair of indices (i,j). The nearest-neighbors of (i,j) are
(i+1,j) in ${\bf a_1}$, and (i,j+1) in ${\bf a_2}$ direction. The Heisenberg
Hamiltonian containing nearest-neighbor interactions defined on this lattice
has $J_1$ coupling in ${\bf a_1}$, and $J_2$ coupling in ${\bf a_2}$
direction. Hence
\begin{eqnarray}
\hat H = \sum_{(i,j)} [(J^x_1 \hat S^x_{i,j} \hat S^x_{i+1,j} + J^y_1
\hat S^y_{i,j} \hat S^y_{i+1,j} + J^z_1 \hat S^z_{i,j} \hat S^z_{i+1,j}) +
(J^x_2 \hat S^x_{i,j} \hat S^x_{i,j+1} + J^y_2
\hat S^y_{i,j} \hat S^y_{i,j+1} + J^z_2 \hat S^z_{i,j} \hat S^z_{i,j+1})]. 
\label{NE8}
\end{eqnarray}
The transformation of spin-1/2 operators in $\hat a_{i,j}, \hat b_{i,j}$ hybrid
operators (anticommute on-site, and commute inter-site) is given by
(\ref{NE1}) generalized to 2D
\begin{eqnarray}
&&\hat S^x_{i,j} = \frac{\hat a^{\dagger}_{i,j} + \hat a_{i,j}}{2X} +
\frac{\hat b^{\dagger}_{i,j} + \hat b_{i,j}}{2Y}, \quad \hat S^y_{i,j} =
\frac{\hat a^{\dagger}_{i,j} - \hat a_{i,j}}{2iZ} +
\frac{\hat b^{\dagger}_{i,j} - \hat b_{i,j}}{2iW},
\nonumber\\
&&\hat S^z_{i,j} = - \frac{1}{2} ( \frac{1}{XZ} + \frac{1}{YW} ) +
(\frac{\hat a^{\dagger}_{i,j} \hat a_{i,j}}{XZ} +\frac{\hat b^{\dagger}_{i,j}
  \hat b_{i,j}}{YW}) + \frac{1}{2} (\frac{1}{YZ} - \frac{1}{XW} )
(\hat a^{\dagger}_{i,j}\hat b^{\dagger}_{i,j} + \hat b_{i,j} \hat a_{i,j} ) 
\nonumber\\
&& \hspace*{0.6cm} + \: \: \frac{1}{2} (\frac{1}{YZ}+ 
\frac{1}{XW})(\hat a^{\dagger}_{i,j} \hat b_{i,j} + \hat b^{\dagger}_{i,j}
\hat a_{i,j} ),
\label{NE9}
\end{eqnarray}
where the scalars $X,Y,Z,W$ satisfy $1=\frac{1}{X^2}+\frac{1}{Y^2}$,
$1=\frac{1}{Z^2}+\frac{1}{W^2}$.

Now (\ref{NE8}) can be expressed as follows: Introducing the notations
$J^x_{\mu}=J_{\mu}(1+\Gamma_{\mu})$, $J^y_{\mu}=J_{\mu}(1-\Gamma_{\mu})$, where
$\mu=1,2$, one finds for the in-plane component
$\hat H_{xy}=\hat H_{xy}(1) + \hat H_{xy}(2)$ the expressions
\begin{eqnarray}
&&\hat H_{xy}(1) = \frac{J_1}{4} \sum_{i,j} \{ [ (X^{-2} -Z^{-2})(
\hat a^{\dagger}_{i,j}\hat a^{\dagger}_{i+1,j} + \hat a_{i,j} \hat a_{i+1,j}) +
(X^{-2} + Z^{-2})(\hat a^{\dagger}_{i,j}\hat a_{i+1,j} + \hat a_{i,j}
\hat a^{\dagger}_{i+1,j}) +
\nonumber\\
&&\hspace*{1cm}+ (Y^{-2} - W^{-2})(\hat b^{\dagger}_{i,j} \hat b^{\dagger}_{i+1,j} +
\hat b_{i,j} \hat b_{i+1,j}) + (Y^{-2} + W^{-2})(\hat b^{\dagger}_{i,j}
\hat b_{i+1,j} + \hat b_{i,j} \hat b^{\dagger}_{i+1,j})
\nonumber\\
&&\hspace*{1cm}+ (X^{-1}Y^{-1}-W^{-1}Z^{-1})
(\hat a^{\dagger}_{i,j} \hat b^{\dagger}_{i+1,j} + \hat a_{i,j} \hat b_{i+1,j}) +
(X^{-1}Y^{-1}+W^{-1}Z^{-1})(\hat a^{\dagger}_{i,j} \hat b_{i+1,j} + \hat a_{i,j}
\hat b^{\dagger}_{i+1,j})
\nonumber\\
&&\hspace*{1cm}+ (X^{-1}Y^{-1}-W^{-1}Z^{-1})(\hat b^{\dagger}_{i,j}
\hat a^{\dagger}_{i+1,j} + \hat b_{i,j} \hat a_{i+1,j}) +  (X^{-1}Y^{-1}+W^{-1}Z^{-1})
(\hat b^{\dagger}_{i,j} \hat a_{i+1,j} + \hat b_{i,j} \hat a^{\dagger}_{i+1,j}) ]
\nonumber\\
&&\hspace*{1cm} + \Gamma_1 \: \: [(X^{-2} +Z^{-2})(\hat a^{\dagger}_{i,j}
\hat a^{\dagger}_{i+1,j} + \hat a_{i,j} \hat a_{i+1,j}) + (X^{-2} - Z^{-2})(
\hat a^{\dagger}_{i,j}\hat a_{i+1,j} + \hat a_{i,j} \hat a^{\dagger}_{i+1,j})
\label{NE10}\\
&&\hspace*{1cm}+ (Y^{-2} + W^{-2})(\hat b^{\dagger}_{i,j}
\hat b^{\dagger}_{i+1,j} + \hat b_{i,j} \hat b_{i+1,j}) + (Y^{-2} - W^{-2})(
\hat b^{\dagger}_{i,j}\hat b_{i+1,j} + \hat b_{i,j} \hat b^{\dagger}_{i+1,j})
\nonumber\\
&&\hspace*{1cm}+ (X^{-1}Y^{-1}+W^{-1}Z^{-1})
(\hat a^{\dagger}_{i,j} \hat b^{\dagger}_{i+1,j} + \hat a_{i,j} \hat b_{i+1,j}) +
(X^{-1}Y^{-1}-W^{-1}Z^{-1})(\hat a^{\dagger}_{i,j} \hat b_{i+1,j} + \hat a_{i,j}
\hat b^{\dagger}_{i+1,j})
\nonumber\\
&&\hspace*{1cm}+ (X^{-1}Y^{-1}+W^{-1}Z^{-1})(\hat b^{\dagger}_{i,j}
\hat a^{\dagger}_{i+1,j} + \hat b_{i,j} \hat a_{i+1,j}) +  (X^{-1}Y^{-1}-W^{-1}Z^{-1})
(\hat b^{\dagger}_{i,j} \hat a_{i+1,j} + \hat b_{i,j} \hat a^{\dagger}_{i+1,j}) ] \}.
\nonumber
\end{eqnarray}
\begin{eqnarray}
&&\hat H_{xy}(2) = \frac{J_2}{4} \sum_{i,j} \{ [ (X^{-2} -Z^{-2})(
\hat a^{\dagger}_{i,j}\hat a^{\dagger}_{i,j+1} + \hat a_{i,j} \hat a_{i,j+1}) +
(X^{-2} + Z^{-2})(\hat a^{\dagger}_{i,j}\hat a_{i,j+1} + \hat a_{i,j}
\hat a^{\dagger}_{i,j+1}) +
\nonumber\\
&&\hspace*{1cm}+ (Y^{-2} - W^{-2})(\hat b^{\dagger}_{i,j} \hat b^{\dagger}_{i,j+1} +
\hat b_{i,j} \hat b_{i,j+1}) + (Y^{-2} + W^{-2})(\hat b^{\dagger}_{i,j}
\hat b_{i,j+1} + \hat b_{i,j} \hat b^{\dagger}_{i,j+1})
\nonumber\\
&&\hspace*{1cm}+ (X^{-1}Y^{-1}-W^{-1}Z^{-1})
(\hat a^{\dagger}_{i,j} \hat b^{\dagger}_{i,j+1} + \hat a_{i,j} \hat b_{i,j+1}) +
(X^{-1}Y^{-1}+W^{-1}Z^{-1})(\hat a^{\dagger}_{i,j} \hat b_{i,j+1} + \hat a_{i,j}
\hat b^{\dagger}_{i,j+1})
\nonumber\\
&&\hspace*{1cm}+ (X^{-1}Y^{-1}-W^{-1}Z^{-1})(\hat b^{\dagger}_{i,j}
\hat a^{\dagger}_{i,j+1} + \hat b_{i,j} \hat a_{i,j+1}) +  (X^{-1}Y^{-1}+W^{-1}Z^{-1})
(\hat b^{\dagger}_{i,j} \hat a_{i,j+1} + \hat b_{i,j} \hat a^{\dagger}_{i,j+1}) ]
\nonumber\\
&&\hspace*{1cm} + \Gamma_2 \: \: [(X^{-2} +Z^{-2})(\hat a^{\dagger}_{i,j}
\hat a^{\dagger}_{i,j+1} + \hat a_{i,j} \hat a_{i,j+1}) + (X^{-2} - Z^{-2})(
\hat a^{\dagger}_{i,j}\hat a_{i,j+1} + \hat a_{i,j} \hat a^{\dagger}_{i,j+1})
\label{NE11}\\
&&\hspace*{1cm}+ (Y^{-2} + W^{-2})(\hat b^{\dagger}_{i,j}
\hat b^{\dagger}_{i,j+1} + \hat b_{i,j} \hat b_{i,j+1}) + (Y^{-2} - W^{-2})(
\hat b^{\dagger}_{i,j}\hat b_{i,j+1} + \hat b_{i,j} \hat b^{\dagger}_{i,j+1})
\nonumber\\
&&\hspace*{1cm}+ (X^{-1}Y^{-1}+W^{-1}Z^{-1})
(\hat a^{\dagger}_{i,j} \hat b^{\dagger}_{i,j+1} + \hat a_{i,j} \hat b_{i,j+1}) +
(X^{-1}Y^{-1}-W^{-1}Z^{-1})(\hat a^{\dagger}_{i,j} \hat b_{i,j+1} + \hat a_{i,j}
\hat b^{\dagger}_{i,j+1})
\nonumber\\
&&\hspace*{1cm}+ (X^{-1}Y^{-1}+W^{-1}Z^{-1})(\hat b^{\dagger}_{i,j}
\hat a^{\dagger}_{i,j+1} + \hat b_{i,j} \hat a_{i,j+1}) +  (X^{-1}Y^{-1}-W^{-1}Z^{-1})
(\hat b^{\dagger}_{i,j} \hat a_{i,j+1} + \hat b_{i,j} \hat a^{\dagger}_{i,j+1}) ] \}.
\nonumber
\end{eqnarray}
Similarly, for the z component in (\ref{NE8}) one obtains
\begin{eqnarray}
\hat H_z = \sum_{\nu=1,2} \sum_n J^z_{\nu} \{ \hat P_1(n,n+\nu) +
\hat P_2(n,n+\nu) + \hat P_3(n,n+\nu) \} = \sum_{\nu=1,2} \hat H_z (\nu)
\label{NE12}
\end{eqnarray}
In this expression the meaning of the abbreviations is $n=(i,j)$,
$n+(\nu=1) = (i+1,j)$, $n+(\nu=2)=(i,j+1)$. One has
\begin{eqnarray}
\hat P_1(n,n+\nu) &=& \frac{1}{4}(\frac{1}{XZ}+\frac{1}{YW})^2 - \frac{1}{2}(
\frac{1}{XZ}+\frac{1}{YW}) \{ [ \frac{\hat a^{\dagger}_n\hat a_n}{XZ} +
\frac{\hat b^{\dagger}_n\hat b_n}{YW} + \frac{1}{2} (\frac{1}{YZ} -\frac{1}{XW})
(\hat a^{\dagger}_n \hat b^{\dagger}_n + \hat b_n \hat a_n)
\nonumber\\
&+& \frac{1}{2} (
\frac{1}{YZ} +\frac{1}{XW})(\hat a^{\dagger}_n \hat b_n + \hat b^{\dagger}_n
\hat a_n) ] +  [ \frac{\hat a^{\dagger}_{n+\nu}\hat a_{n+\nu}}{XZ} +
\frac{\hat b^{\dagger}_{n+\nu}\hat b_{n+\nu}}{YW} + \frac{1}{2} (\frac{1}{YZ} -
\frac{1}{XW})
\nonumber\\
&*&(\hat a^{\dagger}_{n+\nu} \hat b^{\dagger}_{n+\nu} + \hat b_{n+\nu} \hat a_{n+\nu})
+ \frac{1}{2} (\frac{1}{YZ} +\frac{1}{XW})(\hat a^{\dagger}_{n+\nu} \hat b_{n+\nu} +
\hat b^{\dagger}_{n+\nu}\hat a_{n+\nu}) ] \},  
\label{NE13}
\end{eqnarray}
\begin{eqnarray}
\hat P_2(n,n+\nu) &=& (\frac{\hat a^{\dagger}_n\hat a_n}{XZ} +
\frac{\hat b^{\dagger}_n\hat b_n}{YW})(\frac{\hat a^{\dagger}_{n+\nu}\hat a_{n+\nu}}{
XZ} + \frac{\hat b^{\dagger}_{n+\nu}\hat b_{n+\nu}}{YW}) + (\frac{
\hat a^{\dagger}_n\hat a_n}{XZ} +\frac{\hat b^{\dagger}_n\hat b_n}{YW}) [
\frac{1}{2} (\frac{1}{YZ} -\frac{1}{XW})
\nonumber\\
&*&(\hat a^{\dagger}_{n+\nu}\hat b^{\dagger}_{n+\nu} + \hat b_{n+\nu}\hat a_{n+\nu})
+ \frac{1}{2} (\frac{1}{YZ} + \frac{1}{XW})(\hat a^{\dagger}_{n+\nu}
\hat b_{n+\nu} + \hat b^{\dagger}_{n+\nu}\hat a_{n+\nu})]
\nonumber\\
&+& (\frac{
\hat a^{\dagger}_{n+\nu}\hat a_{n+\nu}}{XZ} +\frac{\hat b^{\dagger}_{n+\nu}
\hat b_{n+\nu}}{YW}) [\frac{1}{2}(\frac{1}{YZ} -\frac{1}{XW})(
\hat a^{\dagger}_{n} \hat b^{\dagger}_{n} + \hat b_{n}\hat a_{n})
\nonumber\\
&+& \frac{1}{2} (\frac{1}{YZ} + \frac{1}{XW})(\hat a^{\dagger}_{n}
\hat b_{n} + \hat b^{\dagger}_{n}\hat a_{n})],
\label{NE14}
\end{eqnarray}
\begin{eqnarray}
\hat P_3(n,n+\nu) &=& \frac{1}{4}(\hat a^{\dagger}_n \hat b^{\dagger}_n + \hat b_n
\hat a_n) [ (\frac{1}{YZ}-\frac{1}{XW})^2 (\hat a^{\dagger}_{n+\nu}
\hat b^{\dagger}_{n+\nu} + \hat b_{n+\nu}\hat a_{n+\nu}) + (\frac{1}{(YZ)^2}-
\frac{1}{(XW)^2})
\nonumber\\
&*& (\hat a^{\dagger}_{n+\nu}\hat b_{n+\nu} + \hat b^{\dagger}_{n+\nu}
\hat a_{n+\nu})] + \frac{1}{4}(\hat a^{\dagger}_n \hat b_n + \hat b^{\dagger}_n
\hat a_n) [ (\frac{1}{(YZ)^2}-\frac{1}{(XW)^2}) (\hat a^{\dagger}_{n+\nu}
\hat b^{\dagger}_{n+\nu} + \hat b_{n+\nu}\hat a_{n+\nu})
\nonumber\\
&+& (\frac{1}{YZ}+
\frac{1}{XW})^2 (\hat a^{\dagger}_{n+\nu}\hat b_{n+\nu} + \hat b^{\dagger}_{n+\nu}
\hat a_{n+\nu})].
\label{NE15}
\end{eqnarray}
Finally, one has for the Heisenberg Hamiltonian from (\ref{NE8}), the relation
$\hat H = \hat H_{xy}+\hat H_z$.
\begin{figure}[h]
\includegraphics[height=6cm, width=8cm]{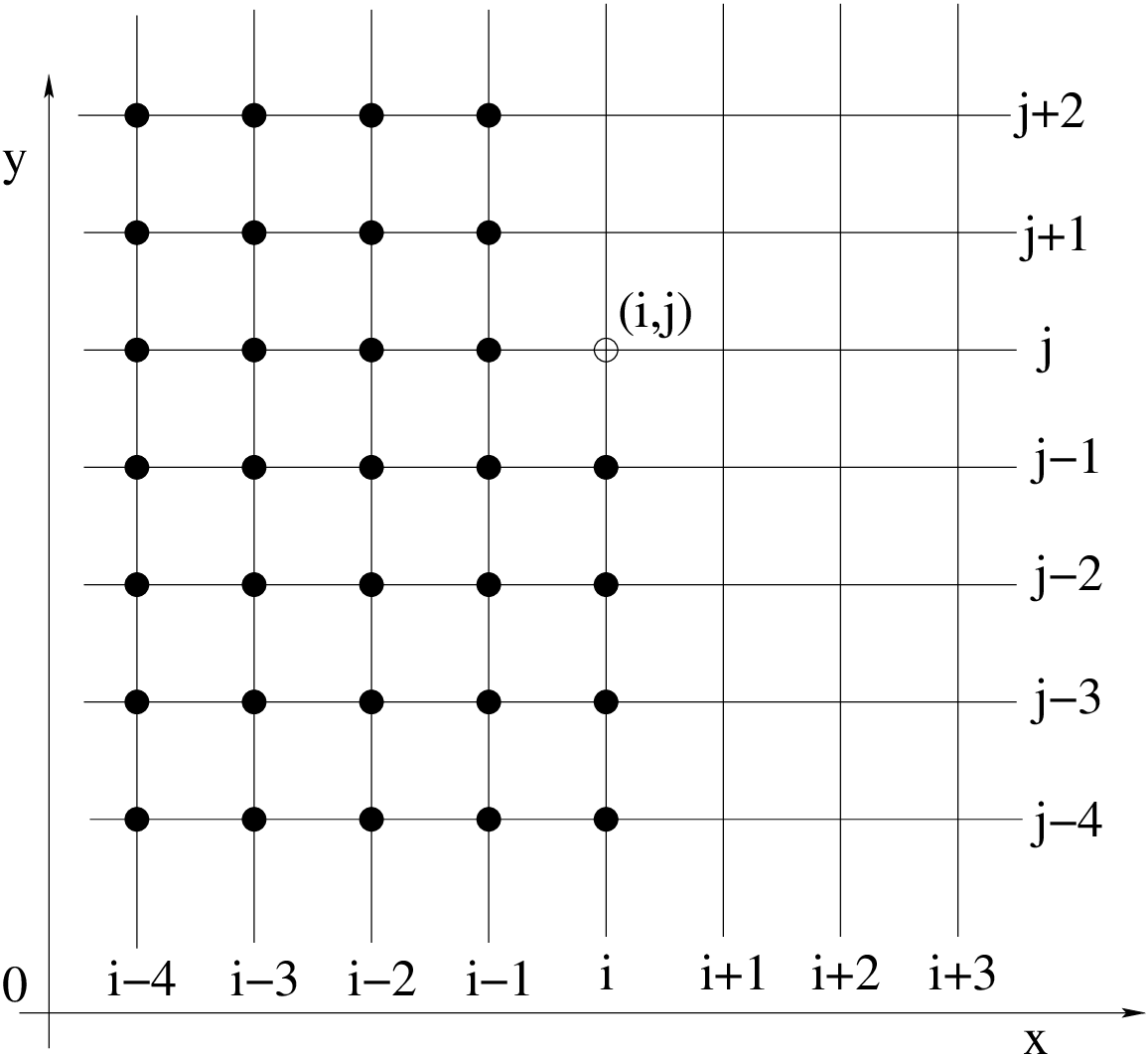}
\caption{Contributions to $\alpha_{i,j}$ in Eqs.(\ref{NE16},\ref{NE17}).
  The studied $(i,j)$ lattice site
  is denoted by an open circle. Black dots are denoting lattice sites
  contributing to $\alpha_{i,j}$.}
\end{figure}

Now we must transform the hybrid $\hat a_{i,j}, \hat b_{i,j}$ operators in
genuine
canonical Fermi operators. This is done via the extension of (\ref{NE2}) to
two dimensions by the expressions
\begin{eqnarray}
&&  \hat a_{i,j}= exp [-i(\hat \alpha^c_{i,j} + \hat \alpha^f_{i,j})]
\hat c_{i,j}, \quad
\hat b_{i,j}= exp [-i(\hat \alpha^c_{i,j} + \hat \alpha^f_{i,j})] \hat f_{i,j},
\nonumber\\
&&  \hat a^{\dagger}_{i,j}= \hat c^{\dagger}_{i,j} exp [+i(\hat \alpha^c_{i,j} +
\hat \alpha^f_{i,j})], \quad
\hat b^{\dagger}_{i,j}= \hat f^{\dagger}_{i,j} exp [+i(\hat \alpha^c_{i,j} +
\hat \alpha^f_{i,j})],  
\label{NE16}
\end{eqnarray}
where $\hat c_{i,j}, \hat f_{i,j}$ are canonical Fermi operators, and
\begin{eqnarray}
\hat \alpha^c_{i,j}= \pi (\sum^{i-1}_{i'=0}\sum^{\infty}_{j'=0} \hat n^c_{i',j'} +
\sum^{j-1}_{j'=0} \hat n^c_{i,j'}), \quad
\hat \alpha^f_{i,j}= \pi (\sum^{i-1}_{i'=0}\sum^{\infty}_{j'=0} \hat n^f_{i',j'} +
\sum^{j-1}_{j'=0} \hat n^f_{i,j'}),
\label{NE17}
\end{eqnarray}
where $\hat n^{c}_{i,j}= \hat c^{\dagger}_{i,j} \hat c_{i,j}$, and
$\hat n^{f}_{i,j}= \hat f^{\dagger}_{i,j} \hat f_{i,j}$.
I note that the 2D extension of (\ref{NE2}) presented in (\ref{NE16},
\ref{NE17}), generalizes for the spinful case the technique used in Ref.
\cite{NR2}.

In order to understand (\ref{NE17}), see Fig.1. I note that for a site (i,j)
(presented as an open circle in Fig.1), in $\hat  \alpha^{\mu}_{i,j}$, $\mu=c,f$,
all $\hat n^{\mu}$ contributions present on the sites denoted by black dots in
Fig.1 are taken into account.

In using (\ref{NE16},\ref{NE17}), first one starts from the canonical Fermi
operator nature of the $\hat c_{i,j}, \hat f_{i,j}$ operators. Given by this,
all $\hat n^{\mu}_{i,j}$ operators, independent on $\mu=c,f$, and $(i,j)$
commute.
Hence, on-site, in bi-quadratic relations, $\hat a_{i,j}$ transform to
$\hat c_{i,j}$, and $\hat b_{i,j}$ to $\hat f_{i,j}$ for all sites (i,j)
(e.g. $\hat a^{\dagger}_{i,j}\hat b_{i,j}=\hat c^{\dagger}_{i,j}\hat f_{i,j}$, etc).
Consequently, the anticommutation nature of
$\hat c_{i,j}, \hat f_{i,j}$ operators imply the local anticommutation of the
$\hat a_{i,j}, \hat b_{i,j}$ operators. On this line I also note, that evaluating
based on (\ref{NE16},\ref{NE17}) e.g. $\{\hat a^{\dagger}_{i,j},
\hat b^{\dagger}_{i,j}
\}=\{\hat c^{\dagger}_{i,j},\hat f^{\dagger}_{i,j}\} $, one uses as well the
relation $\exp[\pm 2i\pi \hat n^{\mu}_{i,j}]=1$ since in the fermionic case, the
particle number operator $\hat n^{\mu}_{i,j}$ has only eigenvalues zero or one.

Turning now on off-site relations, first one observes that in the transformed
expression of the Hamiltonian (\ref{NE10}-\ref{NE15}), two type of off-site
relations are present: a) those connecting the sites $(i,j)$ to $(i,j+1)$, and
b) those connecting the sites $(i,j)$ to $(i+1,j)$. In evaluating these, one
uses the following relations that hold for an arbitrary canonical Fermi
operator $\hat c$, and $\hat n=\hat c^{\dagger} \hat c$, where one also has
$exp[\pm i\pi \hat n ] =1-2 \hat n$:
\begin{eqnarray}
\hat c^{\dagger} (exp[\pm i \pi \hat n]) = \hat c^{\dagger}, \:
(exp[\pm i \pi \hat n]) \hat c^{\dagger} = - \hat c^{\dagger}, \:
\hat c (exp[\pm i \pi \hat n]) = - \hat c, \:
(exp[\pm i \pi \hat n]) \hat c = \hat c.
\label{NE18}
\end{eqnarray}

Let us now start with the case a) mentioned above [(i,j) and (i,j+1) sites
together]. Based on the above provided
information (see (\ref{NE16}-\ref{NE18})), one obtains e.g.
\begin{eqnarray}
&&\hat a^{\dagger}_{i,j} \hat a_{i,j+1} = \hat c^{\dagger}_{i,j} \hat c_{i,j+1} e^{-i\pi
\hat n^f_{i,j}}, \quad \hat a_{i,j+1} \hat a^{\dagger}_{i,j} = - \hat c_{i,j+1}
\hat c^{\dagger}_{i,j} e^{-i\pi \hat n^f_{i,j}},
\nonumber\\
&&\hat b^{\dagger}_{i,j} \hat b_{i,j+1} = \hat f^{\dagger}_{i,j} \hat f_{i,j+1} e^{-i\pi
\hat n^c_{i,j}}, \quad \hat b_{i,j+1} \hat b^{\dagger}_{i,j} = - \hat f_{i,j+1}
\hat f^{\dagger}_{i,j} e^{-i\pi \hat n^c_{i,j}},
\nonumber\\
&&\hat a^{\dagger}_{i,j} \hat b_{i,j+1} = \hat c^{\dagger}_{i,j} \hat f_{i,j+1} e^{-i\pi
\hat n^f_{i,j}}, \quad \hat b_{i,j+1} \hat a^{\dagger}_{i,j} = - \hat f_{i,j+1}
\hat c^{\dagger}_{i,j} e^{-i\pi \hat n^f_{i,j}},
\nonumber\\
&&\hat b^{\dagger}_{i,j} \hat a_{i,j+1} = \hat f^{\dagger}_{i,j} \hat c_{i,j+1} e^{-i\pi
\hat n^c_{i,j}}, \quad \hat a_{i,j+1} \hat b^{\dagger}_{i,j} = - \hat c_{i,j+1}
\hat f^{\dagger}_{i,j} e^{-i\pi \hat n^c_{i,j}},
\nonumber\\
&&\hat a^{\dagger}_{i,j} \hat b^{\dagger}_{i,j+1} = \hat c^{\dagger}_{i,j}
\hat f^{\dagger}_{i,j+1} e^{+i\pi\hat n^f_{i,j}}, \quad \hat b^{\dagger}_{i,j+1}
\hat a^{\dagger}_{i,j} = - \hat f^{\dagger}_{i,j+1}
\hat c^{\dagger}_{i,j} e^{+i\pi \hat n^f_{i,j}},
\nonumber\\
&&\hat b^{\dagger}_{i,j} \hat a^{\dagger}_{i,j+1} = \hat f^{\dagger}_{i,j}
\hat c^{\dagger}_{i,j+1} e^{+i\pi\hat n^c_{i,j}}, \quad \hat a^{\dagger}_{i,j+1}
\hat b^{\dagger}_{i,j} = - \hat c^{\dagger}_{i,j+1}
\hat f^{\dagger}_{i,j} e^{+i\pi \hat n^c_{i,j}},
\nonumber\\
&& --- etc. ---
\label{NE19} 
\end{eqnarray}
Based on (\ref{NE19}), it can be observed that $[\hat a^{\dagger}_{i,j},
\hat a_{i,j+1}] = \{\hat c^{\dagger}_{i,j},\hat c_{i,j+1}\} exp[-i\pi
\hat n^f_{i,j}]$,
$[\hat b^{\dagger}_{i,j},\hat b_{i,j+1}] = \{\hat f^{\dagger}_{i,j},\hat f_{i,j+1}\}
exp[-i\pi \hat n^c_{i,j}]$, etc., hence the anticommutation of the canonical
Fermi operators $\hat c_{i,j}, \hat f_{i',j'}$ automatically implies the off-site
commutation of the $\hat a_{i,j}, \hat b_{i',j'}$ operators.

Furthermore, one observes that in (\ref{NE19}), turning from the bi-quadratic
product of the
hybrid operators $\hat a_{i,j}, \hat b_{i',j'}$ to the bi-quadratic product of the
canonical Fermi operators $\hat c_{i,j}, \hat f_{i',j'}$, a phase factor term
appears (multiplying the canonical Fermi operators), which in the most general
case can be denoted by $\phi^{\mu,\mu'}_{i,j;i',j'}$, where $(i,j)$ denotes the
starting site, $(i',j')$ the ending site, $\mu$ being the annihilation
operator type present at $(i,j)$ (creation operator is denoted by $\bar \mu$),
while $\mu'$ specifies the operator type at $(i',j')$. For example, based on
(\ref{NE19}), one has e.g.
\begin{eqnarray}
\phi^{\bar c,c}_{i,j;i,j+1}=-\pi \hat n^f_{i,j}, \: \:  
\phi^{\bar f,f}_{i,j;i,j+1}=-\pi \hat n^c_{i,j}, \: \:
\phi^{\bar c,f}_{i,j;i,j+1}=-\pi \hat n^f_{i,j}, \: \:
\phi^{\bar f,c}_{i,j;i,j+1}=-\pi \hat n^c_{i,j}, \: \:
\label{NE20}
\end{eqnarray}
Consequently, the transformations from (\ref{NE19}) can be similarly written
as e.g.
\begin{eqnarray}
\hat a^{\dagger}_{i,j} \hat a_{i,j+1} = \hat c^{\dagger}_{i,j} \hat c_{i,j+1}
e^{i\phi^{\bar c,c}_{i,j;i,j+1}}, \: \:
\hat b^{\dagger}_{i,j} \hat b_{i,j+1} = \hat f^{\dagger}_{i,j} \hat f_{i,j+1}
e^{i\phi^{\bar f,f}_{i,j;i,j+1}}, \: \:
\hat a^{\dagger}_{i,j} \hat b_{i,j+1} = \hat c^{\dagger}_{i,j} \hat f_{i,j+1}
e^{i\phi^{\bar c,f}_{i,j;i,j+1}}
\label{NE21}
\end{eqnarray}
Interchanging the operator order in (\ref{NE21}), the phase factor $e^{i\phi}$
remains the same, but given by (\ref{NE18}), a global sign change appears. In
this manner, starting from (\ref{NE16},\ref{NE17}) the hybrid operators
$\hat a_{i,j}, \hat b_{i',j'}$ transform in genuine canonical Fermi operators
$\hat c_{i,j}, \hat f_{i',j'}$.

Note that based on (\ref{NE16},\ref{NE17}) this property remains valid for
arbitrary $(i,j)$ to $(i,j+n)$ term, only the phase factor is changing.
For example
\begin{eqnarray}
&&  \hat a^{\dagger}_{i,j} \hat a_{i,j+n}= \hat c^{\dagger}_{i,j} \hat c_{i,j+n}
exp[i\phi^{\bar c,c}_{i,j;i,j+n}], \quad \hat a_{i,j+n} \hat a^{\dagger}_{i,j}=
-\hat c_{i,j+n} \hat c^{\dagger}_{i,j} exp[i\phi^{\bar c,c}_{i,j;i,j+n}],
\nonumber\\
&& \phi^{\bar c,c}_{i,j;i,j+n} = -\pi[\hat n^f_{i,j} + \sum^{n-1}_{k=1}(
\hat n^c_{i,j+k} + \hat n^f_{i,j+k})]  
\label{NE22}
\end{eqnarray}
Consequently, $[\hat a^{\dagger}_{i,j},\hat a_{i,j+n}] = \{\hat c^{\dagger}_{i,j},
\hat c_{i,j+n}\} exp[i\phi^{\bar c,c}_{i,j;i,j+n}]$, hence, since in the right side
$\{\hat c^{\dagger}_{i,j},\hat c_{i,j+n}\}=0$ holds, $[\hat a^{\dagger}_{i,j},
\hat a_{i,j+n}]=0$ is also satisfied.
Exemplified by Eq.(\ref{NE22}), (see also Eqs.(\ref{NE25},
\ref{NE29})), it can be seen that the presented
transformation can be applied not only to a Heisenberg Hamiltonian as
presented in Eq.(\ref{NE8}), but for an arbitrary spin Hamiltonian written in
terms of quantum spin-1/2 operators.

Let us now analyze the case b) when sites $(i,j)$ and $(i+1,j)$ 
occur together in the transformed Hamiltonian (\ref{NE10}-\ref{NE15}). The
transformations from $\hat a_{i,j}, \hat b_{i',j'}$ to $\hat c_{i,j},
\hat f_{i',j'}$ based on (\ref{NE16},\ref{NE17}) remain basically of the form
presented in (\ref{NE21}), but comparing to (\ref{NE20}) where one component
phases are present, the phase terms $\phi$ become much more complicated even
for nearest neighbors. Let us analyze some examples
\begin{eqnarray}
&&a^{\dagger}_{i,j} \hat a_{i+1,j} = \hat c^{\dagger}_{i,j} \hat c_{i+1,j}
\exp[i\phi^{\bar c,c}_{i,j;i+1,j}], \:
\hat a_{i+1,j} \hat a^{\dagger}_{i,j} = - \hat c_{i+1,j} \hat c^{\dagger}_{i,j}
\exp[i\phi^{\bar c,c}_{i,j;i+1,j}],
\nonumber\\
&&b^{\dagger}_{i,j} \hat b_{i+1,j} = \hat f^{\dagger}_{i,j} \hat f_{i+1,j}
\exp[i\phi^{\bar f,f}_{i,j;i+1,j}], \:
\hat b_{i+1,j} \hat b^{\dagger}_{i,j} = - \hat f_{i+1,j} \hat f^{\dagger}_{i,j}
\exp[i\phi^{\bar f,f}_{i,j;i+1,j}],
\nonumber\\
&&a^{\dagger}_{i,j} \hat b_{i+1,j} = \hat c^{\dagger}_{i,j} \hat f_{i+1,j}
\exp[i\phi^{\bar c,f}_{i,j;i+1,j}], \:
\hat b_{i+1,j} \hat a^{\dagger}_{i,j} = - \hat f_{i+1,j} \hat c^{\dagger}_{i,j}
\exp[i\phi^{\bar c,f}_{i,j;i+1,j}],
\nonumber\\
&&b^{\dagger}_{i,j} \hat a_{i+1,j} = \hat f^{\dagger}_{i,j} \hat c_{i+1,j}
\exp[i\phi^{\bar f,c}_{i,j;i+1,j}], \:
\hat a_{i+1,j} \hat b^{\dagger}_{i,j} = - \hat c_{i+1,j} \hat f^{\dagger}_{i,j}
\exp[i\phi^{\bar f,c}_{i,j;i+1,j}],
\nonumber\\
&&a^{\dagger}_{i,j} \hat b^{\dagger}_{i+1,j} = \hat c^{\dagger}_{i,j} \hat f^{\dagger}_{
i+1,j} \exp[i\phi^{\bar c,\bar f}_{i,j;i+1,j}], \:
\hat b^{\dagger}_{i+1,j} \hat a^{\dagger}_{i,j} = - \hat f^{\dagger}_{i+1,j}
\hat c^{\dagger}_{i,j} \exp[i\phi^{\bar c,\bar f}_{i,j;i+1,j}],
\label{NE23}
\end{eqnarray}
As seen, the transformation to genuine canonic Fermi operators works well also
in this case, since e.g. $[\hat a^{\dagger}_{i,j},\hat a_{i+1,j}] =
\{\hat c^{\dagger}_{i,j},\hat c_{i+1,j}\} exp[i\phi^{\bar c,c}_{i,j;i+1,j}]$, hence,
since in the right side $\{\hat c^{\dagger}_{i,j},\hat c_{i+1,j}\}=0$ holds,
$[\hat a^{\dagger}_{i,j}, \hat a_{i+1,j}]=0$ is also satisfied. The phase factors
however are much more complicated, namely
\begin{eqnarray}
&& \hat \phi^{\bar c,c}_{i,j;i+1,j}=-\pi [\hat n^f_{i,j} + \sum_{\mu=c,f} (
\sum^{\infty}_{j'=j+1} \hat n^{\mu}_{i,j'} + \sum^{j-1}_{j'=0} \hat n^{\mu}_{i+1,j'})],
\nonumber\\
&& \hat \phi^{\bar f,f}_{i,j;i+1,j}=-\pi [\hat n^c_{i,j} + \sum_{\mu=c,f} (
\sum^{\infty}_{j'=j+1} \hat n^{\mu}_{i,j'} + \sum^{j-1}_{j'=0} \hat n^{\mu}_{i+1,j'})],
\nonumber\\
&& \hat \phi^{\bar c,f}_{i,j;i+1,j}=-\pi [\hat n^f_{i,j} + \sum_{\mu=c,f} (
\sum^{\infty}_{j'=j+1} \hat n^{\mu}_{i,j'} + \sum^{j-1}_{j'=0} \hat n^{\mu}_{i+1,j'})],
\nonumber\\
&& \hat \phi^{\bar f,c}_{i,j;i+1,j}=-\pi [\hat n^c_{i,j} + \sum_{\mu=c,f} (
\sum^{\infty}_{j'=j+1} \hat n^{\mu}_{i,j'} + \sum^{j-1}_{j'=0} \hat n^{\mu}_{i+1,j'})],
\nonumber\\
&& \hat \phi^{\bar c,\bar f}_{i,j;i+1,j}= + \pi [\hat n^f_{i,j} + \sum_{\mu=c,f} (
\sum^{\infty}_{j'=j+1} \hat n^{\mu}_{i,j'} + \sum^{j-1}_{j'=0} \hat n^{\mu}_{i+1,j'})],
\label{NE24}
\end{eqnarray}
For better understanding, the sites contributing in the sums present in the
right side of (\ref{NE24}) are presented by black dots in Fig.2.

If at fixed j, instead of the (i+1,j) site (i+n,j) site appears, the
(\ref{NE23}) relations type remains true, e.g.
\begin{eqnarray}
&&a^{\dagger}_{i,j} \hat a_{i+n,j} = \hat c^{\dagger}_{i,j} \hat c_{i+n,j}
\exp[i\phi^{\bar c,c}_{i,j;i+n,j}], \:
\hat a_{i+n,j} \hat a^{\dagger}_{i,j} = - \hat c_{i+n,j} \hat c^{\dagger}_{i,j}
\exp[i\phi^{\bar c,c}_{i,j;i+n,j}],
\nonumber\\
&&b^{\dagger}_{i,j} \hat b_{i+n,j} = \hat f^{\dagger}_{i,j} \hat f_{i+n,j}
\exp[i\phi^{\bar f,f}_{i,j;i+n,j}], \:
\hat b_{i+n,j} \hat b^{\dagger}_{i,j} = - \hat f_{i+n,j} \hat f^{\dagger}_{i,j}
\exp[i\phi^{\bar f,f}_{i,j;i+n,j}],
\label{NE25}
\end{eqnarray}
but in the phase terms (\ref{NE24}), all intermediate columns in between
(i,j) and (i+n,j) enter in the contributing sums
\begin{eqnarray}
&& \hat \phi^{\bar c,c}_{i,j;i+n,j}=-\pi [\hat n^f_{i,j} + \sum_{\mu=c,f} (
\sum^{\infty}_{j'=j+1} \hat n^{\mu}_{i,j'} + \sum^{i+n-1}_{k=i+1}\sum^{\infty}_{j'=0}
 \hat n^{\mu}_{k,j'} +   \sum^{j-1}_{j'=0} \hat n^{\mu}_{i+n,j'})],
\nonumber\\
&& \hat \phi^{\bar f,f}_{i,j;i+n,j}=-\pi [\hat n^c_{i,j} + \sum_{\mu=c,f} (
\sum^{\infty}_{j'=j+1} \hat n^{\mu}_{i,j'} + \sum^{i+n-1}_{k=i+1}\sum^{\infty}_{j'=0}
\hat n^{\mu}_{k,j'} + \sum^{j-1}_{j'=0} \hat n^{\mu}_{i+n,j'})],
\label{NE26}
\end{eqnarray}
\begin{figure}[h]
\includegraphics[height=8cm, width=5cm]{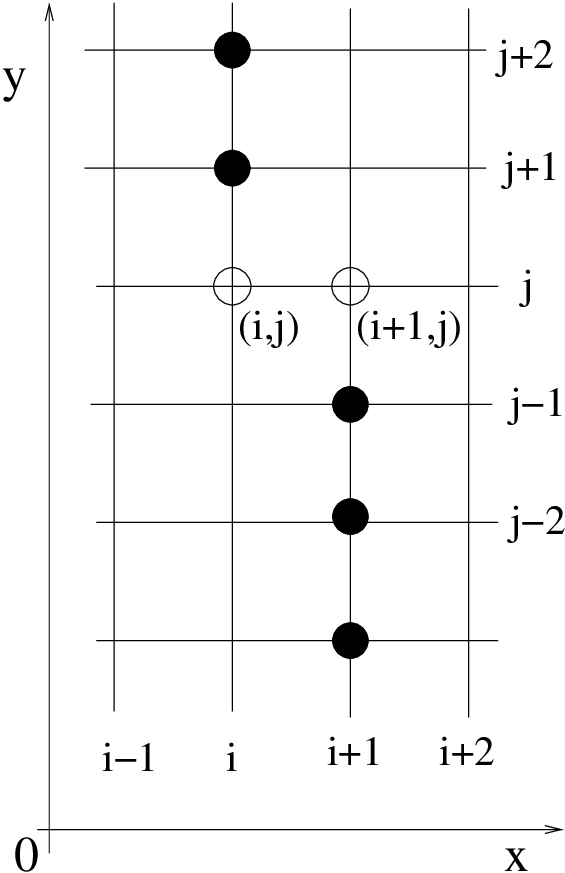}
\caption{Sites contributing in the sum present in $\hat \phi$ phases in
  (\ref{NE24}) are denoted by black dots. The open circles are representing
  the sites (i,j) and (i+1,j).}
\end{figure}
For exemplification, the sites contributing in the sums from (\ref{NE26}) are
presented for the n=2 case in Fig.3a as black dots. Besides, based on
(\ref{NE25}) type of relations, the transformation of the hybrid operators in
canonical Fermi operators is successfully performed at arbitrary n.

Now one considers the most general arbitrary case relating the (i,j) site to
an arbitrary (i+n,j+m) site. The same properties as presented above remain.
For exemplification
\begin{eqnarray}
&&a^{\dagger}_{i,j} \hat a_{i+n,j+m} = \hat c^{\dagger}_{i,j} \hat c_{i+n,j+m}
\exp[i\phi^{\bar c,c}_{i,j;i+n,j+m}], \:
\hat a_{i+n,j+m} \hat a^{\dagger}_{i,j} = - \hat c_{i+n,j+m} \hat c^{\dagger}_{i,j}
\exp[i\phi^{\bar c,c}_{i,j;i+n,j+m}],
\nonumber\\
&&b^{\dagger}_{i,j} \hat b_{i+n,j+m} = \hat f^{\dagger}_{i,j} \hat f_{i+n,j+m}
\exp[i\phi^{\bar f,f}_{i,j;i+n,j+m}], \:
\hat b_{i+n,j+m} \hat b^{\dagger}_{i,j} = - \hat f_{i+n,j+m} \hat f^{\dagger}_{i,j}
\exp[i\phi^{\bar f,f}_{i,j;i+n,j+m}],
\label{NE27}
\end{eqnarray}
\begin{figure}[h]
\includegraphics[height=8cm, width=13cm]{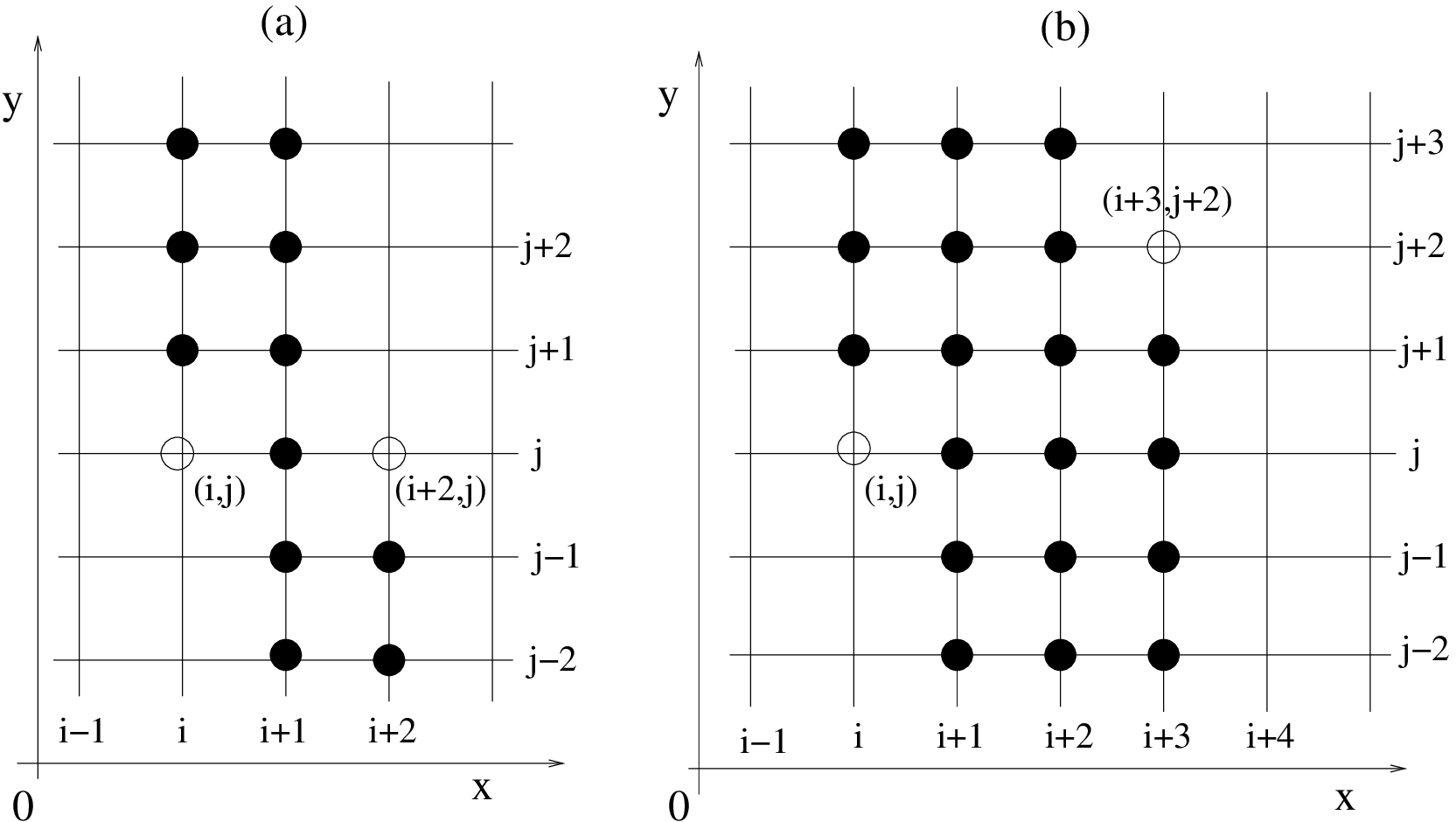}
\caption{a) Sites contributing in the sum present in $\hat \phi$ phases in
  (\ref{NE26}) denoted by black dots in the particular case $n=2$. Note that the
  sites (i,j) and (i+2,j) are denoted by open circles.
  b) Sites contributing in the sum present in $\hat \phi$ phases in
  (\ref{NE28}) denoted by black dots in the particular case $n=3$, $m=2$.
  Note that the open circles are
  representing the sites (i,j) and (i+3,j+2).}
\end{figure}
where one has
\begin{eqnarray}
&& \hat \phi^{\bar c,c}_{i,j;i+n,j+m}=-\pi [\hat n^f_{i,j} + \sum_{\mu=c,f} (
\sum^{\infty}_{j'=j+1} \hat n^{\mu}_{i,j'} + \sum^{i+n-1}_{k=i+1}\sum^{\infty}_{j'=0}
 \hat n^{\mu}_{k,j'} +   \sum^{j+m-1}_{j'=0} \hat n^{\mu}_{i+n,j'})],
\nonumber\\
&& \hat \phi^{\bar f,f}_{i,j;i+n,j+m}=-\pi [\hat n^c_{i,j} + \sum_{\mu=c,f} (
\sum^{\infty}_{j'=j+1} \hat n^{\mu}_{i,j'} + \sum^{i+n-1}_{k=i+1}\sum^{\infty}_{j'=0}
\hat n^{\mu}_{k,j'} + \sum^{j+m-1}_{j'=0} \hat n^{\mu}_{i+n,j'})],
\label{NE28}
\end{eqnarray}
For better understanding, Fig.3b presents the sites contributing in the sums
from (\ref{NE28}) by black dots (in the particular case n=3, m=2). Given by
(\ref{NE27}), as exemplified in (\ref{NE29})
\begin{eqnarray}
&&[\hat a^{\dagger}_{i,j},\hat a_{i+n,j+m}] =
\{\hat c^{\dagger}_{i,j},\hat c_{i+n,j+m}\} exp[i\phi^{\bar c,c}_{i,j;i+n,j+m}],
\nonumber\\
&&[\hat b^{\dagger}_{i,j},\hat b_{i+n,j+m}] =
\{\hat f^{\dagger}_{i,j},\hat f_{i+n,j+m}\} exp[i\phi^{\bar f,f}_{i,j;i+n,j+m}].
\label{NE29}
\end{eqnarray}
As can be seen, the canonical Fermi operator nature of the $\hat c_n, \hat f_n$
operators, based on (\ref{NE16},\ref{NE17}), besides the on-site
anticommutativity, provide also the off-site commutativity property for the
$\hat a_n, \hat b_n$ operators at arbitrary sites. Hence, based on
(\ref{NE16},\ref{NE17}), the transformation in genuine canonical Fermi operators
has been effectuated. Now for these, one takes \cite{NR1}
\begin{eqnarray}
\hat c_{i,j} = \hat c_{i,j,\sigma}, \quad \hat f_{i,j} = \hat c_{i,j,-\sigma}.
\label{NE30}
\end{eqnarray}
Consequently, the transformation in 2D of S=1/2 spin operators to genuine
spinful canonical Fermi operators has been performed.
 
\section{Transcription of the Hamiltonian}

The $\hat H_z$ term (\ref{NE12}-\ref{NE15}) contains only on-site terms, so its
expression in terms of spinful canonical Fermi operators will not contain phase
terms. Concerning the notations, here one has $n=(i,j)$, $n+(\nu=1) = (i+1,j)$,
$n+(\nu=2)=(i,j+1)$. Hence one obtains
\begin{eqnarray}
\hat H_z &=& \sum_{\nu=1,2} \sum_n J^z_{\nu} \{ \hat P_1(n,n+\nu) +
\hat P_2(n,n+\nu) + \hat P_3(n,n+\nu) \},
\nonumber\\  
\hat P_1(n,n+\nu) &=& \frac{1}{4}(\frac{1}{XZ}+\frac{1}{YW})^2 - \frac{1}{2}(
\frac{1}{XZ}+\frac{1}{YW}) \{ [ \frac{\hat c^{\dagger}_{n,\uparrow}
\hat c_{n,\uparrow}}{XZ} + \frac{\hat c^{\dagger}_{n,\downarrow}\hat c_{n,\downarrow}}{
YW} + \frac{1}{2} (\frac{1}{YZ} -\frac{1}{XW})
\nonumber\\
&*&(\hat c^{\dagger}_{n,\uparrow} \hat c^{\dagger}_{n,\downarrow} + \hat c_{n,\downarrow}
\hat c_{n,\uparrow}) + \frac{1}{2} (
\frac{1}{YZ} +\frac{1}{XW})(\hat c^{\dagger}_{n,\uparrow} \hat c_{n,\downarrow} +
\hat c^{\dagger}_{n,\downarrow} \hat c_{n,\uparrow}) ] +  [
  \frac{\hat c^{\dagger}_{n+\nu,\uparrow}\hat c_{n+\nu,\uparrow}}{XZ}
\nonumber\\
&+&\frac{\hat c^{\dagger}_{n+\nu,\downarrow}\hat c_{n+\nu,\downarrow}}{YW}
+ \frac{1}{2} (\frac{1}{YZ} - \frac{1}{XW})
(\hat c^{\dagger}_{n+\nu,\uparrow} \hat c^{\dagger}_{n+\nu,\downarrow} +
\hat c_{n+\nu,\downarrow} \hat c_{n+\nu,\uparrow})
+ \frac{1}{2} (\frac{1}{YZ} +\frac{1}{XW})
\nonumber\\
&*&(\hat c^{\dagger}_{n+\nu,\uparrow}
\hat c_{n+\nu,\downarrow} +\hat c^{\dagger}_{n+\nu,\downarrow}\hat c_{n+\nu,\uparrow})
] \},  
\label{NE31}
\end{eqnarray}
\begin{eqnarray}
\hat P_2(n,n+\nu) &=& (\frac{\hat c^{\dagger}_{n,\uparrow}\hat c_{n,\uparrow}}{XZ} +
\frac{\hat c^{\dagger}_{n,\downarrow}\hat c_{n,\downarrow}}{YW})(\frac{
 \hat c^{\dagger}_{n+\nu,\uparrow}\hat c_{n+\nu,\uparrow}}{XZ} + \frac{
 \hat c^{\dagger}_{n+\nu,\downarrow}\hat c_{n+\nu,\downarrow}}{YW}) + (\frac{
  \hat c^{\dagger}_{n,\uparrow}\hat c_{n,\uparrow}}{XZ} +\frac{
  \hat c^{\dagger}_{n,\downarrow}\hat c_{n,\downarrow}}{YW}) [
\frac{1}{2} (\frac{1}{YZ} -\frac{1}{XW})
\nonumber\\
&*&(\hat c^{\dagger}_{n+\nu,\uparrow}\hat c^{\dagger}_{n+\nu,\downarrow} +
\hat c_{n+\nu,\downarrow}\hat c_{n+\nu,\uparrow})
+ \frac{1}{2} (\frac{1}{YZ} + \frac{1}{XW})(\hat c^{\dagger}_{n+\nu,\uparrow}
\hat c_{n+\nu,\downarrow} + \hat c^{\dagger}_{n+\nu,\downarrow}\hat c_{n+\nu,\uparrow})]
\nonumber\\
&+& (\frac{
\hat c^{\dagger}_{n+\nu,\uparrow}\hat c_{n+\nu,\uparrow}}{XZ} +\frac{
\hat c^{\dagger}_{n+\nu,\downarrow} \hat c_{n+\nu,\downarrow}}{YW})
[\frac{1}{2}(\frac{1}{YZ} -\frac{1}{XW})(
\hat c^{\dagger}_{n,\uparrow} \hat c^{\dagger}_{n,\downarrow} + \hat c_{n,\downarrow}
\hat c_{n,\uparrow})
\nonumber\\
&+& \frac{1}{2} (\frac{1}{YZ} + \frac{1}{XW})(\hat c^{\dagger}_{n,\uparrow}
\hat c_{n,\downarrow} + \hat c^{\dagger}_{n,\downarrow}\hat c_{n,\uparrow})],
\label{NE32}
\end{eqnarray}
\begin{eqnarray}
\hat P_3(n,n+\nu) &=& \frac{1}{4}(\hat c^{\dagger}_{n,\uparrow}
\hat c^{\dagger}_{n,\downarrow} + \hat c_{n,\downarrow} \hat c_{n,\uparrow}) [
(\frac{1}{YZ}-\frac{1}{XW})^2 (\hat c^{\dagger}_{n+\nu,\uparrow}
\hat c^{\dagger}_{n+\nu,\downarrow} + \hat c_{n+\nu,\downarrow}\hat c_{n+\nu,\uparrow}) +
(\frac{1}{(YZ)^2}-\frac{1}{(XW)^2})
\nonumber\\
&*& (\hat c^{\dagger}_{n+\nu,\uparrow}\hat c_{n+\nu,\downarrow} +
\hat c^{\dagger}_{n+\nu,\downarrow} \hat c_{n+\nu,\uparrow})] + \frac{1}{4}(
\hat c^{\dagger}_{n,\uparrow} \hat c_{n,\downarrow} + \hat c^{\dagger}_{n,\downarrow}
\hat c_{n,\uparrow}) [ (\frac{1}{(YZ)^2}-\frac{1}{(XW)^2})
\nonumber\\
&*&(\hat c^{\dagger}_{n+\nu,\uparrow} \hat c^{\dagger}_{n+\nu,\downarrow} +
\hat c_{n+\nu,\downarrow}\hat c_{n+\nu,\uparrow})
+ (\frac{1}{YZ}+
\frac{1}{XW})^2 (\hat c^{\dagger}_{n+\nu,\uparrow}\hat c_{n+\nu,\downarrow} +
\hat c^{\dagger}_{n+\nu,\downarrow} \hat c_{n+\nu,\uparrow})].
\label{NE33}
\end{eqnarray}

The $\hat H_{xy}(2)$ term has phase factors containing only one term (see
(\ref{NE20},\ref{NE21})), so this term also can be easily written. One obtains
\begin{eqnarray}
\hat H_{xy}(2) &=& \sum_{i,j} \sum_{\sigma} [ K_{2,\sigma} \hat c^{\dagger}_{i,j;\sigma}
\hat c^{\dagger}_{i,j+1;\sigma}(1-2 \hat n_{i,j;-\sigma}) + t_{2,\sigma}
\hat c^{\dagger}_{i,j;\sigma} \hat c_{i,j+1,\sigma}(1-2 \hat n_{i,j;-\sigma})
\nonumber\\
&+& K_2
\hat c^{\dagger}_{i,j;\sigma}\hat c^{\dagger}_{i,j+1;-\sigma}(1-2 \hat n_{i,j;-\sigma})
+ t_2 \hat c^{\dagger}_{i,j;\sigma} \hat c_{i,j+1,-\sigma}(1-2 \hat n_{i,j;-\sigma}) +
h.c.],
\label{NE34}
\end{eqnarray}
where, using the coupling constants of the Hamiltonian (\ref{NE8}), we
define $J^x_{\mu}=J_{\mu}(1+\Gamma_{\mu})$, $J^y_{\mu}=J_{\mu}(1-\Gamma_{\mu})$, where
$\mu=1,2$. With these notations one obtains the coupling constants from
(\ref{NE34}) using $\mu=2$ in the relations (\ref{NE35}) presented below:
\begin{eqnarray}
&& K_{\mu,\sigma} = \frac{J_{\mu}}{4} [ (\frac{1}{X^2}-\frac{1}{Z^2}) + \Gamma_{\mu}
(\frac{1}{X^2}+\frac{1}{Z^2})],
\nonumber\\
&& K_{\mu,-\sigma} = \frac{J_{\mu}}{4} [ (\frac{1}{Y^2}-\frac{1}{W^2}) + \Gamma_{\mu}
(\frac{1}{Y^2}+\frac{1}{W^2})] = \frac{J_{\mu}}{4} [-(\frac{1}{X^2}-\frac{1}{
Z^2}) + \Gamma_{\mu} (2-(\frac{1}{X^2}+\frac{1}{Z^2}))],
\nonumber\\
&& t_{\mu,\sigma} = \frac{J_{\mu}}{4} [ (\frac{1}{X^2}+\frac{1}{Z^2}) + \Gamma_{\mu}
(\frac{1}{X^2}-\frac{1}{Z^2})],
\nonumber\\ 
&& t_{\mu,-\sigma} = \frac{J_{\mu}}{4} [ (\frac{1}{Y^2}+\frac{1}{W^2}) + \Gamma_{\mu}
(\frac{1}{Y^2}-\frac{1}{W^2})] =
\frac{J_{\mu}}{4} [(2- (\frac{1}{X^2}+\frac{1}{Z^2}) - \Gamma_{\mu}
(\frac{1}{X^2}-\frac{1}{Z^2})],
\nonumber\\
&& K_{\mu} = \frac{J_{\mu}}{4} [ (\frac{1}{XY}-\frac{1}{WZ}) + \Gamma_{\mu}
(\frac{1}{XY}+\frac{1}{WZ})] = \frac{J_{\mu}}{4} [ A + \Gamma_{\mu} B],
\nonumber\\
&& t_{\mu} = \frac{J_{\mu}}{4} [ (\frac{1}{XY}+\frac{1}{WZ}) + \Gamma_{\mu}
(\frac{1}{XY}-\frac{1}{WZ})] = \frac{J_2}{4} [ B + \Gamma_{\mu} A].
\label{NE35}
\end{eqnarray}
Taking into account that only the $X,Z$ parameters are independent, one has
in (\ref{NE35}) $AB= \frac{1}{X^2}(1- \frac{1}{X^2}) -\frac{1}{Z^2}(1-
\frac{1}{Z^2})$, $A= \frac{1}{XY}-\frac{1}{WZ}$, and $B=AB/A$. Note that
$\frac{1}{Y^2}=1-\frac{1}{X^2}$ and $\frac{1}{W^2}=1-\frac{1}{Z^2}$ hold.

The remaining $\hat H_{xy}(1)$ term is the genuine land of the phase factor
contributions (see (\ref{NE23},\ref{NE24})). Here, in order to write down
explicitly the emerging terms, the sum part of the phases from (\ref{NE24})
will be separately denoted by
\begin{eqnarray}
\hat I_{i,j} = exp[-i\pi \sum_{\mu=c,f} (
\sum^{\infty}_{j'=j+1} \hat n^{\mu}_{i,j'} + \sum^{j-1}_{j'=0} \hat n^{\mu}_{i+1,j'})]
\label{NE36}
\end{eqnarray}
When applied to a wave vector, both $\hat I_{i,j}$, and $\hat I^*_{i,j}$ provide
the same sign +1 or -1. Furthermore, since all terms in $\hat H_{x,y}(1)$
are multiplied by $\hat I_{i,j}$, or $\hat I^*_{i,j}$, $\hat I_{i,j}$ it can be
extracted from
all terms under the sum in (\ref{NE10}). The remaining phase factors in
(\ref{NE24}) coincide with the phase factors from (\ref{NE20}). Consequently
$\hat H_{xy}(1)$ will maintain the form of (\ref{NE34}), with the own coupling
constants, and the $\hat I_{i,j}$ term under the sum. Consequently
\begin{eqnarray}
\hat H_{xy}(1) &=& \sum_{i,j} \hat I_{i,j} \sum_{\sigma} [ K_{1,\sigma}
\hat c^{\dagger}_{i,j;\sigma} \hat c^{\dagger}_{i+1,j;\sigma}(1-2 \hat n_{i,j;-\sigma})
+ t_{1,\sigma} \hat c^{\dagger}_{i,j;\sigma} \hat c_{i+1,j,\sigma}(1-2 \hat n_{i,j;-
\sigma})
\nonumber\\
&+& K_1 \hat c^{\dagger}_{i,j;\sigma}\hat c^{\dagger}_{i+1,j;-\sigma}(1-2
\hat n_{i,j;-\sigma})
+ t_1 \hat c^{\dagger}_{i,j;\sigma} \hat c_{i+1,j,-\sigma}(1-2 \hat n_{i,j;-\sigma}) +
h.c.],
\label{NE37}
\end{eqnarray}
where the coupling constants are obtained from (\ref{NE35}) using the $\mu=1$
value. As a consequence $\hat H_{x,y} = \hat H_{xy}(1) + \hat H_{x,y}(2)$, and the
Heisenberg Hamiltonian (\ref{NE8}) written in term of spinful canonical Fermi
operators becomes $\hat H = \hat H_z + \hat H_{x,y}$ based on (\ref{NE32}-
\ref{NE37}). The fact that $\hat I_{i,j}$ is site dependent shows why the 1D
Heisenberg Hamiltonian is much more easy to treat than the 2D Heisenberg case.

For completeness I present also  the expression of the Zeeman term
$\hat H_{Zeem} = - h \sum_n S^z_n$
\begin{eqnarray}
\hat H_{Zeem} &=& -h \sum_n \{ [ (\frac{\hat c^{\dagger}_{n,\uparrow}
\hat c_{n,\uparrow}}{XZ} +\frac{\hat c^{\dagger}_{n,\downarrow}
\hat c_{n,\downarrow}}{YW}) -\frac{1}{2} ( \frac{1}{XZ} + \frac{1}{YW} )]
+ \frac{1}{2} (\frac{1}{YZ} - \frac{1}{XW} )
(\hat c^{\dagger}_{n,\uparrow}\hat c^{\dagger}_{n,\downarrow} + \hat c_{n,\downarrow}
\hat c_{n,\uparrow} ) 
\nonumber\\
&+& \: \: \frac{1}{2} (\frac{1}{YZ}+ 
\frac{1}{XW})(\hat c^{\dagger}_{n,\uparrow} \hat c_{n,\downarrow} +
\hat c^{\dagger}_{n,\downarrow} \hat c_{n,\uparrow} ) \}.
\label{NE38}
\end{eqnarray}

\section{Summary and conclusions}

The paper presents the generalization to two dimensions of the recently
published 1D generalized Jordan-Wigner transformation connecting the S=1/2
quantum spin operators to genuine spinful canonical Fermi operators \cite{NR1}.

I note that several extensions from 1D to 2D are known for the Jordan-Wigner
transformation connecting S=1/2 spin operators to spinless fermion operators.
These in essence differ
by the phase terms emerging during the transformation procedure. This is because
\cite{NR2,NR3,NR4} these phases appearing on the fermionic side of the
transformation, can be considered to be related to a gauge field created by the
spin system itself. From these, I was choosing and generalizing for the
treatment of the spinful fermionic case analyzed here the technique from
(\cite{NR2}), because this allows to write explicitly the transformed spin
Hamiltonian in terms of the spinful canonical Fermi operators, since it provides
explicitly the phase values emerging during the transformation. Since the
transformation is exact, this gives the possibility to observe what kind of
characteristics at Heisenberg spin system level correspond to given
characteristics at fermionic level, and then to establish mappings in between
these two systems. On this line e.g. one observes that xy
spin-spin interaction in a given direction in the spin system correspond to
(at most) correlated movement (hopping) in that direction at fermionic level
(so it is not surprising, that the 1D $\hat H_{xy}$ spin model is solved by an
1D itinerant and non-interacting fermionic system \cite{NR6}). Or z
spin-spin coupling in a given direction at the spin system level provides 
nearest-neighbor Coulomb interaction in that direction at the fermionic level.
Furthermore, the $\hat H_{xy}(1)$ term also provides interesting information:
i) The emerging $\hat I_{i,j}$ term, emphasizes why the 2D Heisenberg model is
much more difficult to treat in comparison to the 1D Heisenberg case. ii) This
$\hat I_{i,j}$ factor (containing infinite many contributions in the
thermodynamic limit) introduces also long-range contributions in the 2D
fermionic model describing the 2D local Heisenberg system. This underlines the
observation at the point i). iii) Only  $\hat H_{xy}(1)$ contains
the $\hat I_{i,j}$ contribution (in the presented transformation). At the
fermionic level it looks like (mimic, it seems to be equivalent to) an
unidirectional disordered contribution (as an unidirectional random hopping).
Considering $J^x_1,J^y_1$ small, this could be used as a perturbation. This
perturbative treatment could supply supplementary information about the 2D
Heisenberg system besides the mean-field type of approaches usually used
\cite{NR2,NR3}.

I would like to present also mapping possibilities. The technique from
Ref.(\cite{NR3}) suggests that all sites are equally touched by similarly
cumbersome phase factors. But using the technique from Ref.(\cite{NR2}) one
observes that in fact the influence of the cumbersome phase factors can be
collected on a given group of site-pairs (here those contained in
$\hat H_{xy}(1)$). Hence it is possible to obtain exact relatively workable
mappings
between the 2D Heisenberg spin system, and 2D spinful Fermi systems avoiding
these groups. An example is presented below emerging at $J_2^z=\Gamma_2=0$,
$X=Z, Y=W$ obtaining on the spinful fermionic side
\begin{eqnarray}
 \hat H &=& \sum_{i,j} \{ \sum_{\sigma} [t_{2,\sigma} \hat c^{\dagger}_{i,j;\sigma}
 \hat c_{i,j+1;\sigma}(1-2 \hat n_{i,j;-\sigma}) +  t_{2} \hat c^{\dagger}_{i,j;\sigma}
 \hat c_{i,j+1;-\sigma}(1-2 \hat n_{i,j;-\sigma}) + h.c.]
\nonumber\\
&+& Q [\hat c^{\dagger}_{i,j;
 \uparrow}  \hat c_{i,j;\downarrow}(\frac{\hat n_{i+1,j;\uparrow}}{X^2} +  
\frac{\hat n_{i+1,j;\downarrow}}{Y^2} -\frac{1}{2}) + \hat c^{\dagger}_{i+1,j;
 \uparrow}  \hat c_{i+1,j;\downarrow}(\frac{\hat n_{i,j;\uparrow}}{X^2} +  
\frac{\hat n_{i,j;\downarrow}}{Y^2} -\frac{1}{2}) + h.c]
\nonumber\\
&+& V_1 (\hat c^{\dagger}_{
 i+1,j; \uparrow}  \hat c_{i+1,j;\downarrow} +h.c.)(\hat c^{\dagger}_{i,j; \uparrow}
 \hat c_{i,j;\downarrow} +h.c.) +V_2(\frac{\hat n_{i+1,j;\uparrow}}{X^2} +  
\frac{\hat n_{i+1,j;\downarrow}}{Y^2})(\frac{\hat n_{i,j;\uparrow}}{X^2} +  
\frac{\hat n_{i,j;\downarrow}}{Y^2})
\nonumber\\
&+& V_2[\frac{1}{4} -( \frac{
\hat n_{i+1,j;\uparrow}+ \hat n_{i,j;\uparrow}}{2X^2} +  
\frac{\hat n_{i+1,j;\downarrow}+\hat n_{i,j;\downarrow}}{2Y^2})] \},
\label{NE39}
\end{eqnarray}
whose image on the 2D Heisenberg side of the transformation is
\begin{eqnarray}
\hat H = \sum_{(i,j)} [J^z_1 \hat S^z_{i,j} \hat S^z_{i+1,j} +
J_2 (\hat S^x_{i,j} \hat S^x_{i,j+1} + \hat S^y_{i,j} \hat S^y_{i,j+1})]. 
\label{NE40}
\end{eqnarray}
For the coupling constants of (\ref{NE39}) one has
\begin{eqnarray}
  t_{2,\uparrow}=\frac{J_2}{2X^2}, \quad t_{2,\downarrow}=\frac{J_2}{2Y^2}, \quad
  t_2=\frac{J_2}{2XY}, \quad Q=\frac{J^z_1}{XY}, \quad V_1 = \frac{J^z_1}{X^2Y^2},
  \quad V_2=J^z_1.
  \label{NE41}
\end{eqnarray}
I exemplify the mapping possibilities by (\ref{NE39}), since this Hamiltonian
describes in fact movement along one direction in a two dimensional system, a
subject which is intensively analyzed nowadays. E.g. effects of the uni-axial
strain in 2D systems \cite{NR7}, target selection in 2D \cite{NR8},
anisotropic hopping in 2D \cite{NR9}, or even
actuators enhancing unidirectional movement used in the production of shape
memory devices \cite{NR10}, fields in which the interconnection in between
(\ref{NE39}-\ref{NE40}) could provide valuable information.

Intensive future work is needed for the clarification, understanding and use
of these mapping possibilities and characteristics. 

%
%
%




\end{document}